\documentstyle[12pt,aasms4,flushrt]{article}
\begin{document}
\received{ }
\accepted{  }
\journalid{ }{ }
\slugcomment{   }
%
%
\lefthead{Barrado y Navascue\'es, Stauffer, \& Patten }
\righthead{The Age of IC~2391}

\title{The Lithium Depletion Boundary and the 
Age of the Young Open Cluster IC~2391 }

\author{David Barrado y Navascu\'es\altaffilmark{1},
\affil{Max-Planck-Institut f\"ur Astronomie. K\"onigstuhl 17,
Heidelberg, D-69117 Germany}
\and
 John R. Stauffer,  Brian M. Patten\altaffilmark{1},
} 
\affil{Harvard--Smithsonian Center for Astrophysics,
       60 Garden St., Cambridge, MA 02138, USA}

\altaffiltext{1}{Visiting Astronomer, 
Cerro Tololo Inter-American Observatory. 
CTIO is operated by AURA, Inc.\ under contract
 to the National Science Foundation.}

\begin{abstract}

In previous programs, we have identified a large number
of possible very low mass members of the young, open
cluster IC~2391 based on their location in an $I$ versus 
($R-I$)$_c$ color-magnitude diagram.  We have now obtained
new photometry and intermediate resolution ($\Delta \lambda = 2.7$ \AA\ )
spectra of 19 of these objects (14.9 $\le$ $I_c$ $\le$ 17.5) in order 
to confirm cluster membership. We identify 15 of our targets as 
likely cluster members based on their $VRI$ photometry, spectral types, 
radial velocity, and H$\alpha$ emission strengths.  Higher S/N
spectra were obtained for 8 of these probable cluster members in 
order to measure the strength of the lithium 6708 \AA\  doublet and 
thus obtain an estimate of the cluster's age. 
One of these 8 stars has a definite lithium  detection and two other 
(fainter) stars have possible lithium detections.  A color-magnitude
diagram for our program objects shows that the lithium depletion
boundary in IC~2391 is at $I_c$=16.2.  Using recent theoretical
model predictions, we derive an age for IC~2391 of
53$\pm$5 Myr.  While this is considerably older than the age most
commonly attributed for this cluster ($\sim$35 Myr) this result
for IC~2391 is comparable those recently derived for the Pleiades and Alpha 
Persei clusters and can be explained by new models for high mass
stars that incorporate a modest amount of convective core overshooting.

\end{abstract}


\section{Introduction}

Measurement of the location of the ``lithium depletion boundary'' (LDB) offers
a new method to determine the age of young open clusters that is claimed
to be more accurate and less subject to possible systematic errors than
any other means (c.f., Bildsten et al. 1997; Basri 1997; 
Ventura et al. 1998a,b;
Ushomirsky et al. 1998).  The LDB method was first applied to the Pleiades 
cluster (Basri, Marcy, \& Graham 1996; Rebolo et al. 1996), with the most 
recent determination (Stauffer, Schultz, \& Kirkpatrick 1998) yielding an age 
of 125$\pm$8 Myr.  That age is considerably older than some estimates of the 
Pleiades age, but consistent with other estimates which assume a modest 
amount of convective core overshoot (Ventura et al. 1998a,b).
Our group has now obtained LDB ages for two other open clusters -- Alpha
Persei (Stauffer et al. 1999 --see also Basri \& Mart\'{\i}n 1999)
 and IC~2391.  We report the results for 
IC~2391 in this paper.

Located at a distance of about $\sim$155 pc, IC~2391 ($\alpha$=8$^h40.2^m$, 
$\delta$=--53$^\circ$04', J2000.0) is the fifth nearest 
cluster to the Sun.  The first faint proper motion survey of this 
cluster was obtained by Stauffer et al. (1989), yielding 9 previously 
undiscovered late-type members.  Patten \& Simon (1996) used data from ROSAT 
to obtain a more complete membership list by combining the X-ray data 
with ground-based optical photometry and spectroscopy.  Additional
spectroscopic studies by Stauffer et al. (1997) confirmed that the cluster 
is relatively young based on rotational velocities, lithium abundances and 
chromospheric activity for G, K and early M dwarfs.   However, none of
those studies provided cluster members extending faint enough ($I > 15$) to 
be suitable for the LDB test.  Recently, Patten \& Pavlovsky (1999) and
Barrado y Navascu\'es et al. (1999a) have provided lists of faint candidate
members of IC~2391 which may extend to low enough mass.  Spectra of
some of these candidates are used here to determine the location of the LDB
in IC~2391.

\section{Observations and Data Reduction}

Spectroscopic observations of the most promising candidate low-mass members 
of the cluster were carried out with the 4m telescope at CTIO
 during 9--11 January 1999. We used the R-C 
spectrograph with the Blue Air Schmidt camera, the KPGLD grating (790 l/mm), 
the Loral 3K CCD, and a one arcsec slit.  This configuration yielded a 
spectral resolution of 2.7 \AA\  and a free spectral range of 6295--8815 \AA.

The primary targets for our observations were selected from the optical 
survey by 
Barrado y Navascu\'es et al. (1999a), which identified several dozen very
low mass stars and brown dwarf candidates based primarily on $R_c$,$I_c$ 
photometry.  The targets were selected to have apparent magnitudes 
near to or below the expected location of the LDB in 
IC~2391 using the nominal age and distance to the cluster and the theoretical 
model predictions of Chabrier \& Baraffe (1997).
Additional $V$ filter data, obtained at the CTIO 0.9m telescope using the
Tektronix 2K \#3 camera during 4--6 January 1999, allowed us to refine the 
target list
based on their location in a [($V-R_c$),($R-I$)$_c$] color-color diagram.
Near infrared data from the 2MASS project (Adams 1998) for some 
of our objects were also used
to refine the target list.  Additional details can be found in
Barrado y  Navascu\'es et al. (1998, 1999a).
Several, presumably higher mass objects, were added to our target list 
from the surveys of Patten \& Pavlovsky (1999) and Patten \& Simon (1996).
We also observed several Gliese stars of spectral types M0--M6 and
BRI~0021-0214, a M9.5 spectral type dwarf (Basri \& Marcy 1995),
in order to serve as template objects with which to compare the IC~2391
spectra. For the IC~2391 low-mass candidates, the exposure times ranged 
from 10 minutes to 4.5 hours.  The data were reduced using standard 
routines within the IRAF\footnote{IRAF is distributed  by National 
Optical Astronomy Observatories, which is operated by
the Association of Universities for Research 
in Astronomy, Inc., under contract to the National
Science Foundation, USA}.

\section{Analysis}

We have determined equivalent widths and line profiles for our
program stars for several spectral features, namely H$\alpha$ in emission, 
\ion{Li}{1} 6708\AA, and the \ion{K}{1} and \ion{Na}{1} doublets 
(7665/7699 and 8183/8195 \AA, respectively).  In addition, we have measured 
several spectral indices, namely PC1, PC2 and PC3 at 6525--7050\AA, 
7030-7580\AA, and 7540-8265\AA, respectively (Mart\'{\i}n, Rebolo, 
\& Zapatero-Osorio 1996); VO at 7350-7560\AA\ (Kirkpatrick, Henry, 
\& Simons 1995); and TiO, CaH at 6510-7050 \AA\ (Mart\'{\i}n \& Kun 1996), 
in order to estimate spectral types and ($R-I$)$_c$\ colors.  These estimates 
were achieved by a calibration based on a group of cool dwarfs of known 
spectral types and colors.  The derived spectral types are estimated to have 
an uncertainty of 0.5 or 1 subclass.  Radial velocity estimates were derived 
for the program stars by two methods: (1) by cross-correlation versus the 
spectra of GJ285 and GJ406 and (2) by measurement of the wavelength of the 
H$\alpha$ emission line.  The radial velocity errors for the cross-correlation
method are estimated to range from 10 to 50 km/s.  The uncertainty in the 
radial velocity derived from the fit to the H$\alpha$ profile should be 
$\sim$15 km/s for most of our objects based on the repeatability in the 
measured position of OH night-sky emission lines of similar strength in the 
sky spectra.  Table 1 lists the names of our targets, along with the 
photometry (Barrado y Navascu\'es et al. 1999a) and the spectroscopic 
results.   

\subsection{Membership Criteria}

In the last column of Table 1, we provide an estimate of the 
membership status for each of our target stars.  This status was determined 
by consideration of the following criteria:
\medskip

\noindent (1) A comparison of the photometrically measured (R-I)$_C$ 
color and the color estimated from the spectral type, assuming
E($R-I$)$_C$ $\sim$ 0.01) for IC~2391 (e.g., Hogg 1960; Patten \& Simon 1996),
was used to eliminate background objects.

\noindent (2) Radial velocities were compared with the cluster average 
of $\sim$15 km/sec (Stauffer et al. 1997).

\noindent (3) H$\alpha$ equivalent widths were compared with typical values
for field stars and Pleiades members.

\noindent (4) The strength of the gravity dependent \ion{Na}{1} 
8200\AA\ doublet,
strong in dM, weaker in PMS M, and very weak in M-type giants (Mart\'{\i}n, 
Rebolo, \& Zapatero-Osorio 1996), was used to eliminate background
giants and nearby field stars (see Figure 1).

These criteria allowed us to reject two stars as members of the cluster
and to classify another two stars as possible non-members.  CTIO-040 and
CTIO-094 have radial velocities less than 3 km/s.  For these two stars, the 
difference in their photometrically measured ($R-I$)$_C$ 
color and the color estimated from the spectral type is 0.17 and 0.24 
magnitudes respectively. Both of these stars also have H$\alpha$ in absorption.
In the spectral type range we have covered, all members of the 
Pleiades have H$\alpha$ in emission with EW$\sim$2--10 \AA\  
(Stauffer, Liebert, \& Giampapa 1995).  Because IC~2391 is younger, we
expect to see similar or stronger H$\alpha$ emission from our candidates.   
Therefore we consider CTIO-040 and CTIO-094 to be nearly certain non-members.  
The targets CTIO-081 and CTIO-106 were found to have radial velocities
less than 3 km/s.  These stars are considered possible non-members.
The remaining 15 targets in our sample appear to be young, PMS objects and 
thus are likely members of the IC 2391 cluster.

\subsection{The Lithium Depletion Boundary in IC~2391}

Figure 2 shows the region around the lithium doublet for three members whose 
$I$ magnitudes place them near the LDB 
-- PP14, CTIO-096, and CTIO-038.  The spectrum of GJ406, an M6V star without 
lithium, is shown for comparison purposes. Clearly, CTIO-038 ($I_c$=16.29) 
contains lithium, whereas PP14, 0.46 magnitudes brighter, does not have a 
discernible feature at 6708\AA. CTIO-096, slightly brighter than CTIO-038, 
has also apparently depleted all of its lithium. We also have low 
signal-to-noise, possible detections of lithium in two other candidate 
members of IC~2391 (CTIO-041 and CTIO-062), both of which are fainter and 
redder than CTIO-038.  Other candidate cluster members, brighter and warmer 
than CTIO-038 and observed at high S/N, do not show lithium in their spectra 
(see Table 1).  Figure 3 shows an  $I_c$/($R-I$)$_c$ color-magnitude diagram 
for our sample of candidates. Based on the detection of 
lithium in CTIO-038 and the non-detection of lithium in CTIO-096, we can 
estimate the location of the LDB in IC~2391 to be 
at $I_c$=16.2$\pm$0.15, ($R-I$)$_c$=1.91.  

Using the evolutionary models of Baraffe et al. (1998), we can estimate the
age of IC~2391 using the position of LDB.
Following Figure 3 of Stauffer, Schultz, \& Kirkpatrick (1998), we have 
converted the $I_c$ magnitude of the LDB into an age, 
assuming an interstellar absorption of A$_{\rm I}$=0.02 and a distance 
modulus of ($m-M$)$_{\rm 0}$=5.95 $\pm$ 0.1\footnote{This value for the 
distance of IC~2391 represents a compromise between various determinations
of the distance modulus and {\it Hipparcos} parallaxes.  See, for example,
Becker \& Fenkart (1971), Lynga 1987, Patten \& Pavlovsky (1999), and 
Robichon et al. (1999) for additional details.}
yielding  $M_{I}$ = 10.25 which implies an age of 53$\pm$5 Myr.
While this value of the cluster's age is significantly larger than previous 
estimates ($\sim$35 Myr), this correction factor to the classical age
is consistent with those estimated for other young open clusters, such 
as the Pleiades (Stauffer, Schultz, \& Kirkpatrick 1998) and $\alpha$ Persei 
(Stauffer et al. 1999), using the LDB method.

\subsection{The Lowest Mass Members of IC~2391}

At an age of 53 Myr, a star at the LDB should have
a mass of $\sim$0.12 M$_\odot$ (Chabrier \& Baraffe 1997).  The corresponding 
$I_c$ magnitude for a 0.075 M$_\odot$ object (i.e., an object at the 
transition boundary between the stellar and sub-stellar domains) in 
IC 2391 is then $M_{I}$ $\sim$ 11.15.  If the latter estimate is correct, then
the faintest two candidate members in IC~2391 that we have observed 
spectroscopically (CTIO-061 and CTIO-113) would be brown dwarfs.

\section{Discussion}

The ages for young open clusters have been controversial for many years
due to uncertainties in the amount of convective core overshoot to
include in models of high mass stars.  Although an age scale for clusters
using models without convective core overshoot has generally been
adopted (Patenaude 1978; Mermilliod 1981), models which
incorporate a relatively large convective core overshoot parameter,
yielding cluster ages up to twice as old as the no-overshoot models, have
also been advocated (Mazzei \& Pigatto 1988).  Without
additional observational data, it has been difficult to choose
between these age scales.  The lithium depletion age method provides
the means to resolve this debate.   We now have lithium depletion
ages for three open clusters: the Pleiades (Stauffer, Schultz, \& Kirkpatrick 
1998), Alpha Persei (Basri \& Mart\'{\i}n 1999; Stauffer et al. 1999), and 
IC~2391 (this paper).  When compared to the non-overshoot model ages of 
Mermilliod (1981), the ratio of the LDB age determination to the 
non-overshoot model age is 1.61, 1.65, and 1.46 for Pleiades, Alpha Persei and 
IC~2391, respectively.   This suggests that (a) some convective-core 
overshoot is needed in evolutionary models for high-mass stars
and (b) the amount of convective-core overshoot is not a strong
function of mass (at least in the mass range sampled at the turn-off
of these three clusters).  A revised age scale for open clusters would
have important implications for a variety of stellar evolution
topics including, for example, the empirical (and theoretical)
correlations between lithium abundance, coronal activity, and rotation
rate as a function of age for low mass stars.

\acknowledgements

DBN thanks the IAC (Spain) and the
DFG (Germany) for their fellowship. JRS acknowledges support from NASA Grant
NAGW-2698 and 3690.

\newpage  

\begin{center}
{\sc Figure Captions}
\end{center}

\figcaption{\ion{Na}{1} 8200 \AA\  equivalent width versus spectral type. 
Field dwarfs are shown as solid circles and IC~2391 members are shown as 
open circles. Stars initially identified as candidate IC~2391 members but 
subsequently rejected as members using the membership criteria outlined 
in \S 3.1 appear as crosses.
\label{fig1}}

\figcaption{Spectra of IC~2391 candidates ({\it solid lines}).
We only show here the spectral region around the \ion{Li}{1} 6708 \AA\  
doublet ({\it vertical dashed line}).  For comparison, GJ406, field M6 dwarf
with no lithium, is also plotted ({\it dotted line}).
\label{fig2}}

\figcaption{Color-magnitude diagram of IC~2391. 
Probable  members observed at low S/N appear as plus symbols ($+$), 
whereas members observed at high S/N are displayed as circles (solid if 
lithium was detected, open if not). Solid triangle symbols indicate probable 
lithium detections. Non-members appear as $\times$ symbols). 
The location of the LDB is indicated, as well as a
 ZAMS (solid line) and 50 and 30 isochrones (dashed lines).
\label{fig3}}

\newpage


\newpage

\begin{thebibliography}{ }

\bibitem[1998]{Adams1998}
Adams, J. 1998, personal communication

\bibitem[1998]{Baraffe1998}
Baraffe, I. 1998, personal communication

\bibitem[1998]{ByN1998}
Barrado y Navascu\'es, D., Stauffer, J. R., \& Brice\~no, C. 1998,
In ``Very Low Mass Stars and Brown Dwarfs  in Stellar Clusters
and Associations'', IAC, Eds., La Palma, Spain, in press

\bibitem[1999]{ByN1999a}
Barrado y Navascu\'es, D., et al. 1999a, in preparation

\bibitem[1999]{ByN1999b}
Barrado y Navascu\'es, D., Stauffer, J. R., Bouvier, J. 1999b Ap\&SS, in press

\bibitem[1997]{Basri1997}
Basri, G.  1997, in Cool Stars in Clusters and 
Associations, ed. R. Pallavicini \& G. Micela, Mem. Soc. Astron. Italiana,
 68 (4), 917

\bibitem[1995]{Basri1995}
Basri, G., \& Marcy, G. W. 1995, AJ 109, 762

\bibitem[1996]{Basri1996}
Basri, G., Marcy, G., \& Graham, J. 1996, ApJ 458, 600

\bibitem[1999]{Basri1999}
Basri, G., \& Mart\'{\i}n, E. L. 1999, ApJ 510, 266

\bibitem[1971]{Becker1971}
Becker, W., \& Fenkart, R. 1971, A\&AS 4, 241

\bibitem[1997]{Bildsten1997}
Bildsten, L., Brown, E., Matzer, C., \& Ushomirski, G. 1997, ApJ 482, 442 

\bibitem[1997]{Chabrier1997}
Chabrier, G., \& Baraffe, I. 1997, A\&A 327, 1039

\bibitem[1994]{DAntona1994}
D'Antona, F, \& Mazzitelli, I. 1994, ApJS 90, 467

\bibitem[1961]{Feinstein1961}
Feinstein, A. 1961, PASP 73, 410

\bibitem[1960]{Hogg1960}
Hogg, A. R. 1960, PASP 72, 85

\bibitem[1996]{Jones1996}
Jones, B. F., Shetrone, M., \& Fischer, D. 1996, AJ 112, 186

\bibitem[1995]{Kirkpatrick1995}
Kirkpatrick, J. D., Henry, T. J., \& Simons, D. A. 1995, AJ 109, 797

\bibitem[1984]{Levato1984}
Levato, H., \& Malaroda, S. 1984, Ap.Lett. 24, 37

\bibitem[1987]{Lynga1987}
Lynga, G. 1987, Catalog of Open Cluster Data (5th Ed., Lund: Lund Observatory)

\bibitem[1996]{Martin1996a}
Mart\'{\i}n, E. L., \& Kun, M. 1996, A\&AS 116, 467

\bibitem[1996]{Martin1996b}
Mart\'{\i}n, E. L., Rebolo, R., \& Zapatero-Osorio, M. R. 1996, ApJ 469, 706

\bibitem[1988]{Mazzei1988}
Mazzei, P., \& Pigatto, L. 1988, A\&A 193, 148

\bibitem[1981]{Mermilliod1981}
Mermilliod, J. C. 1981, A\&A 97, 235

\bibitem[1978]{Patenaude1978}
Patenaude, M. 1978, A\&A 66, 225

\bibitem[1996]{Patten1996} 
Patten, B. M., \& Simon, T. 1996, ApJ 106, 489

\bibitem[1999]{Patten1999} 
Patten, B. M., \& Pavlovski, C. M. 1999, PASP 111, 210

\bibitem[1969]{Perry1969} 
Perry, C. L., \& Hill, G. 1969, AJ 74, 899

\bibitem[1996]{Rebolo1996}
Rebolo, R., Mart\'{\i}n, E. L., Basri, G., \& Zapatero-Osorio, M. R. 1996,
ApJ 469, L53

\bibitem[1999]{Robichon1999}
Robichon, N., Arenou, F., Mermilliod, J.-C., \& Turon, C. 1999, A\&A in press


\bibitem[1989]{Stauffer1989}
Stauffer, J. R., Hartmann, L. W., Jones, B. J., \& McNamara, B. R. 1989, 
ApJ 342, 285

\bibitem[1995]{Stauffer1995}
Stauffer, J. R., Liebert, J., \& Giampapa, M. 1995, AJ 109, 298

\bibitem[1997]{Stauffer1997}
Stauffer, J. R., Hartmann, L. W., Prosser, C. F., Randich, S., 
Patten, B. M., Simon, T., \& Giampapa, M. 1997, ApJ 479, 776

\bibitem[1998]{Stauffer1998}
Stauffer, J. R., Schultz, G., \& Kirkpatrick, J. D. 1998, ApJ 449, L199

\bibitem[1999]{Stauffer1999} 
Stauffer, J. R., Barrado y Navascu\'es, D., Bouvier, J., Morrison, H. L., 
Hardig, P., Luhman, K., Stanke, T., McCaughrean, M., Terndrup, D. M., 
Allen, L., \& Assouad, P. 1999, ApJ, submitted

\bibitem[1998]{Ushomirsky1998}
Ushomirsky, G., Matzner, C. D., Brown, E. F., Bildsten, L., Hilliard, V. G., 
\& Schroeder, P. C. 1998, ApJ 497, 253 

\bibitem[1998]{Ventura1998a}
Ventura, P., Zeppieri, A., Mazzitelli, I., \& D'Antona, F. 1998a, 
A\&A 331, 1011

\bibitem[1998]{Ventura1998b}
Ventura, P., Zeppieri, A., Mazzitelli, I., \& D'Antona, F. 1998b, 
A\&A 334, 953
 
\end{thebibliography}
\end{document}